\documentclass[reprint, superscriptaddress, preprintnumbers, 
amsmath,amssymb, aps, prl,
]{revtex4-2}
\usepackage{graphicx, braket, float}
\usepackage[usenames,dvipsnames,x11names]{xcolor}
\usepackage[english]{babel}
\usepackage[colorlinks=true, citecolor = MidnightBlue, linkcolor = MidnightBlue, urlcolor=MidnightBlue, pdfborder={0 0 0}]{hyperref}
\usepackage{orcidlink}
\usepackage{times}
\usepackage{newtxmath}

\addto\captionsenglish{}
\newcommand{\nth}{n_\text{{th}}^{(\text{in})}}
\newcommand{\an}{\hat{a}_j}
\newcommand{\ad}{\hat{a}^\dagger_j}

\begin{document}

\title{Observation of entanglement in a cold atom analog of cosmological preheating}

\newcommand{\LCF}{Université Paris-Saclay, Institut d’Optique Graduate School, CNRS, Laboratoire Charles Fabry, 91127, Palaiseau, France}
\author{Victor Gondret\,\orcidlink{0009-0005-8468-161X}}
\email{victor.gondret@normalesup.org}
\affiliation{\LCF}
\author{Clothilde Lamirault\,\orcidlink{0009-0001-6468-2181}}
\affiliation{\LCF}
\author{Rui Dias\,\orcidlink{0009-0004-4158-7693}}
\affiliation{\LCF}
\author{Léa Camier\,\orcidlink{0009-0003-4345-3608}}
\affiliation{\LCF}
\author{Amaury Micheli\,\orcidlink{0000-0002-5240-140X}}
\affiliation{RIKEN Center for Interdisciplinary Theoretical and Mathematical Sciences (iTHEMS), Wako, Saitama 351-0198, Japan}
\author{Charlie Leprince\,\orcidlink{0009-0002-5490-6767}}
\affiliation{\LCF}
\author{Quentin Marolleau\,\orcidlink{0009-0002-3587-3912}} \altaffiliation{Present address: Qblox, Delftechpark, Netherlands.}
\affiliation{\LCF}
\author{Jean-Ren\'e Rullier}
\affiliation{\LCF}
\author{Scott Robertson\,\orcidlink{0000-0001-5919-8320}}
\affiliation{Institut Pprime, CNRS -- Université de Poitiers -- ISAE-ENSMA. TSA 51124, 86073 Poitiers Cedex 9, France}
\author{Denis Boiron\,\orcidlink{0000-0002-2719-5931}} \email{denis.boiron@institutoptique.fr}\affiliation{\LCF}
\author{Christoph I. Westbrook\,\orcidlink{0000-0002-6490-0468}} \email{christopher.westbrook@institutoptique.fr}
\affiliation{\LCF}

\begin{abstract}
	\noindent
We observe entanglement between collective excitations of a Bose-Einstein condensate in a configuration analogous to particle production during the preheating phase of the early universe. In our setup, the oscillation of the inflaton field is mimicked by the transverse breathing mode of a cigar-shaped condensate, which parametrically excites longitudinal quasiparticles with opposite momenta. After a short modulation period, we observe entanglement of these pairs which reveals the role played by vacuum fluctuations in seeding the parametric growth, confirming the quantum origin of the excitations. As the system continues to evolve, we observe a decrease in correlations and a disappearance of non-classical features. These point towards future experimental probes of the late-time nonlinear regime where further analogies can be drawn with reheating, i.e. the thermalization of the post-inflationary Universe.
\end{abstract}
\maketitle

Quantum fields are subject to vacuum fluctuations, which can be amplified to create particles. This phenomenon plays a crucial role in astrophysics and cosmology and is known as spontaneous particle production, to distinguish it from its stimulated counterpart seeded by classical fluctuations.
It is the underlying mechanism behind black hole evaporation via Hawking radiation ~\cite{Hawking:1974rv}, the generation of primordial cosmological inhomogeneities during inflation~\cite{Mukhanov:1981xt}, 
and the generation of particles in an empty post-inflationary universe, a phenomenon known as preheating~\cite{Kofman:1997yn,Bassett:2005xm}.
However, a direct observation of these phenomena in the cosmological context is currently out of reach~\cite{Unruh-1981, campo.2006.bell, Micheli:2023qnc}.

Unruh showed~\cite{Unruh-1981} that, in the presence of a strong coherent background, the excitations of a fluid, or \textit{quasi}particles, can be treated using the same formalism as particles in a curved spacetime.
Following this idea, analogs of Hawking radiation~\cite{philbin.2008.fiber, weinfurtner.2011.measurement, steinhauer.2016.observation, euve.2016.observation, observation.2019.drori, munoz_de_nova_observation_2019,falque.2025.polariton}, the cosmological redshift~\cite{eckel.2018.expanding}, quasiparticle production in time-varying geometries~\cite{wilson.2011.observation,hu.2019.quantum, sparn.2024.experimental}, and false-vacuum decay~\cite{zenesini_false_2024} have been realized among others~\cite{barcelo_analogue_2011}.
However, the entanglement between the produced excitations, a signature of their quantum origin, is fragile and elusive, and has only been probed in a few experiments. In the context of time-dependent Hamiltonians, these experiments involved quenched Bose-Einstein condensates (BECs)~\cite{chen.2021.observation, tenart_observation_2021} and photonic experiments with modulated boundary conditions
analogous to the dynamical Casimir effect~\cite{Lahteenmaki:2011cwo, vezzoli.2019.optical}.

We report the observation of entangled, parametrically excited, collective modes in a BEC in an analog to cosmological preheating. 
In the early Universe, a hypothetical, nearly homogeneous field, known as the inflaton, first drives an exponential expansion of the universe (inflation). Subsequently, when oscillating around the minimum of its potential, it provides a periodic drive which parametrically amplifies the vacuum fluctuations of otherwise empty fields leading to particle production (preheating).
In our system, the role of the inflaton is played by the transverse degrees of freedom of an elongated BEC excited in a breathing mode. 
The width of the gas oscillates coherently across the condensate, parametrically producing quasiparticles in the longitudinal degrees of freedom. The temperature is sufficiently low for the growth to be seeded mostly by vacuum fluctuations, resulting in an entangled state. The analogy thus applies only to the preheating phase~\footnote{Our analogy is incomplete as it does not capture the effect of space expansion during preheating. While in our work the resonance proceeds in a single narrow band, space expansion causes the modes to move through the Mathieu instability bands, usually limiting efficient preheating to situations with broad resonance bands~\cite{Kofman:1997yn,Bassett:2005xm}, still see~\cite{Greene:1997fu} for a conformally invariant case where the expansion of space can be factored out.}. 
Though this type of parametric amplification had already been studied in quantum fluids and is often referred to as Faraday wave generation~\cite{engels.2007.faraday, smits.2018.observation, nguyen.2019.parametric, hernandez.2021.faraday, liebster.2025.prx, jaskula.2012.acoustic, clark.2017.collective}, entanglement was not observed~\footnote{Parametric amplification and entanglement generation in quantum gases have also been studied in  spinor condensates~\cite{pezze.2018.RevModPhys} and with single-species BECs using  other excitation schemes such as colliding condensates~\cite{perrin.2007.observation, jaskula.2010.subpoissonian, hodgman.2017.solving}, collisional de-excitation in a waveguide~\cite{bucker.twinatom.2011, borselli.2021.two-particle} or parametric instability in a moving lattice~\cite{campbell.2006.parametric, bonneau.2013.tunable}.}. 
Here, the very low temperature allows us to certify entanglement at low quasiparticle number. As the system continues to evolve, the quasiparticle population grows exponentially and we observe a decrease of the quasiparticle correlation which we relate to a possible loss of entanglement. This decrease indicates a transition to thermalization in further analogy with the reheating stage in cosmology, which follows preheating.


\begin{figure*}
    \includegraphics[width=7.06in]{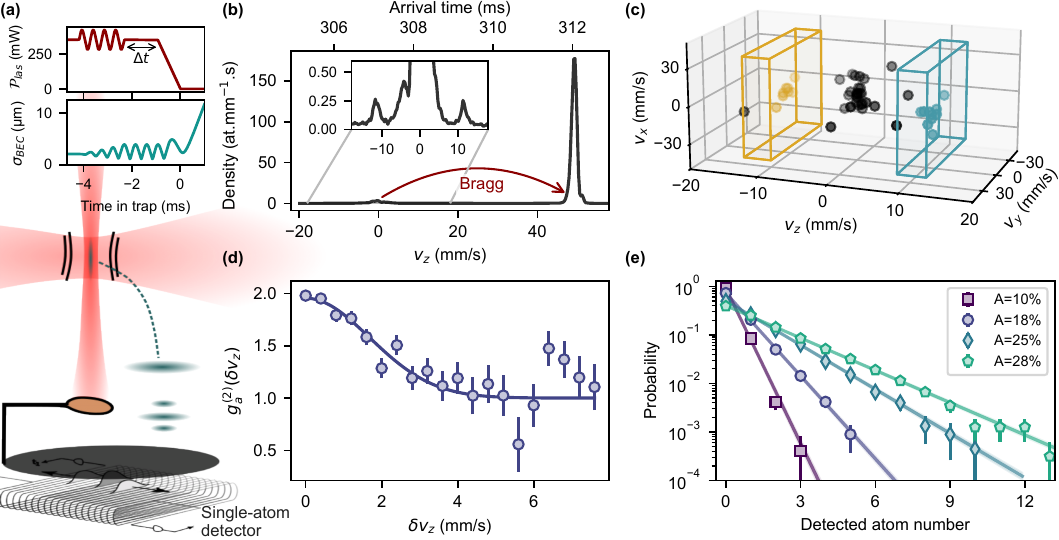}
    \caption{
    (a)~Diagram of the experimental apparatus and excitation protocol (see text for description).
    (b)~The position and arrival time of individual atoms is recorded, and converted to an initial velocity. Sidebands are visible at $\pm 11.7$~mm/s. A Bragg diffraction pulse shifts more than 97\% of the BEC atoms to later times to avoid saturating the detector in the vicinity of the excitations.
    (c)~A single shot showing the excitations in a 3D velocity space. Each dot represents a single atom. The boxes show the position and size of a typical analysis volume or ``voxel''.
    (d)~Auto-correlation function of the measured sideband velocities giving an estimate of the longitudinal mode size and showing its thermal nature. 
    (e)~Probability distribution in a single voxel for different modulation depths $ A$ each of these acquired over $\sim2800$ realizations. 
    The lines show the probability distribution in Eq.~(\ref{eqn:ThermalStateDistribuition}) computed from the mean detected atom numbers of 0.094(6), 0.37(1), 0.99(3) and 1.50(3)  for increasing value of $A$.}
\label{fig1}
\end{figure*}
\textit{Experiment---}A BEC containing about $3\,000$ metastable helium atoms in the $2^3\mathrm{S}_{1}$, $m_S=1$ state is trapped in a crossed optical dipole trap, see Fig.~\ref{fig1}(a) and the End Matter.
The trapping frequencies are $\omega_\perp/2\pi=930(20)$\,Hz in the radial direction and $\omega_{z}/2\pi=40(2)$~Hz in the vertical direction.
In an elongated homogeneous cloud, longitudinal collective excitations with momentum $q$ have an energy $\hbar\omega_q$ given by the Bogoliubov dispersion relation~\cite{pitaevskii2016bose}
\begin{equation}\label{eq:bogoliubov}
\hbar\omega_q = \sqrt{c_{1\text{D}}^2 q^2+\left(\frac{q^2}{2m}\right)^2}
\end{equation}
where $c_{1\text{D}}$ is the longitudinal sound speed, $m$ the atomic mass, and $\hbar$ the reduced Planck constant. 
A periodic modulation of the sound speed parametrically
produces correlated pairs of quasiparticles at momenta $\pm p$, whose frequency $\omega_{\pm p}$ is half the modulation frequency~\cite{engels.2007.faraday,jaskula.2012.acoustic,clark.2017.collective, nguyen.2019.parametric, hernandez.2021.faraday, liebster.2025.prx}.
The dynamics can be modeled as a two-mode squeezing operation on the $\pm p$ quasiparticle modes, which can become entangled 
depending on the initial temperature of the cloud, the thermal damping of the quasiparticles, and the strength of the modulation~\cite{busch.2014.quantum}. 
In a harmonic trap, momentum states are not the exact eigenmodes of the system, but the physics at play is not expected to change significantly~\cite{butera.2021.position}.
At the center of the trap, we estimate the local sound speed to be 7(1)~mm/s \footnote{From the dispersion relation~\eqref{eq:bogoliubov} and the quasiparticle energy $\hbar\omega_\perp$, we estimate the Bogoliubov coefficients associated to the excited quasiparticles to be $1.03(2)$ and $-0.24(8)$, as defined in Eq.~(19) of Ref.~\cite{robertson.2017.controlling}.}.

The periodic modulation of $c_{1\mathrm{D}}$ is achieved by varying the power of the dipole trap $\mathcal{P}_{\text{las}}\propto (1+A\sin 2\omega_\perp t)$. 
This excites the transverse breathing mode of the BEC, whose radius $\sigma_{\mathrm{BEC}}\propto1/c_{1D}$ oscillates at $2\omega_\perp$~\cite{gerbier.1D.2004}. The breathing amplitude depends on the modulation amplitude $A$ and its duration, kept constant at 4 modulation periods in this work. After the modulation, the trap remains on for a hold time $\Delta t$ during which the transverse radius $\sigma_{\text{BEC}}$ continues to oscillate at $2\omega_\perp$, see the inset of Fig.~\ref{fig1}(a).
A peculiar feature of the breathing mode in an elongated cloud is its low damping rate~\cite{chevy.2002.transverse} caused by the suppression of Landau damping mechanisms, as both the thermal cloud and the BEC oscillate in phase~\cite{jackson.accidental.2002}.

After the excitation process, the transverse confinement is ramped down in 1~ms after which the longitudinal confinement is also switched off. 
This ramp is sufficiently slow to ensure that the collective excitations are mapped onto the atoms by a process referred to as phonon evaporation~\cite{tozzo.2004.phonon, robertson.2017.controlling, gondret.2025.parametric}.
During the free fall, these atoms fly away from the BEC, appearing as sidebands in the time-of-flight spectrum relative to the unperturbed condensate when they reach a microchannel plate (MCP) detector, see Fig.~\ref{fig1}(b).
The MCP detector identifies the arrival time and position of individual atoms with a quantum efficiency of 25(5)\%. The detector is located 46~cm below the trap, sufficiently far that the arrival times and positions accurately reflect the atomic velocities when the trap was turned off.

The MCP is shielded from the vertical laser by a metal disc, requiring the atoms to be deflected towards the unshielded region.
This is achieved by a $12$~\textmu s stimulated Raman transition which transfers the atoms to the magnetically insensitive $m_S = 0$ state and imparts a transverse velocity of 42~mm/s.
In addition, to prevent the saturation of the MCP caused by the high atomic flux from the BEC, a temporally shaped, velocity-selective Bragg pulse, applied 1~ms after the Raman pulse, shifts the BEC to later times thus removing 97\% of the BEC atoms from the region where we detect the excitations~\cite{leprince.2025.coherent}.
A typical single shot velocity reconstruction is shown in Fig.~\ref{fig1}(c) where each dot represents an individual atom.
The boxes show the position and size of a typical analysis volume or ``voxel''.

\textit{Single mode statistics --} A typical one-dimensional velocity profile is shown in the inset of Fig.~\ref{fig1}(b). 
A Gaussian fit to the sideband density profiles yields typical standard deviations of $\sigma_\perp \sim 8$~mm/s in the transverse direction and $\sigma_z \sim 0.8$~mm/s longitudinally.
To compare these scales against that of a single mode, we construct a histogram of velocity differences between atom pairs. 
This allows us to compute the normalized two-body auto-correlation function~\cite{schellekens_hanbury_2005}
\begin{equation}
g^{(2)}_{a}(\boldsymbol{\delta v}) = \int_\Omega \frac{\braket{:\hat{n}_{\boldsymbol{v}-\boldsymbol{\delta v}} \hat{n}_{\boldsymbol{v}+\boldsymbol{\delta v}}:}}{\braket{\hat{n}_{\boldsymbol{v}+\boldsymbol{\delta v}}} \braket{\hat{n}_{\boldsymbol{v}-\boldsymbol{\delta v}}}} \text{d}^{3}\boldsymbol{v},
\end{equation}
where $\hat{n}$ is the atom number operator, ``:'' refer to normal ordering and the integration volume $\Omega$ excludes the residual condensate~\footnote{The region $\Omega$ takes into account all atoms within a volume $[-40,-8]\cup[8,40]$~mm/s along $v_z$ and $[-40,40]$~mm/s along $v_x$ and $v_y$. 
The plot shows $g^{(2)}_{a}(\delta v_z)=\iint_{-\Delta_\perp/2}^{\Delta_\perp/2} g^{(2)}(\boldsymbol{\delta v})\text{d}\delta v_{x,y}^2$ with a transverse integration of $\Delta_{\perp}=2\sigma_{\perp}$.}.
The sidebands exhibit bunching within a momentum linewidth inversely proportional to the source size in each spatial direction~\cite{gomes.2006.theory,butera.2021.position}. 
We show in Fig.\ref{fig1}(d) the auto-correlation function $g_a^{(2)}$ along the $z$ direction. 
A Gaussian fit to this peak gives a standard deviation of $1.7(1)$ mm/s, which we interpret as the characteristic momentum width of a single mode~\cite{gomes.2006.theory}. A similar analysis along the transverse direction yields a standard deviation of 9(1)~mm/s.  Comparing these to $\sigma_{i}$, we confirm that our excitation mainly concerns pairs of single modes. 

To probe the statistics of these excitations more precisely, a voxel in momentum space is defined—illustrated in Fig.~\ref{fig1}(c)—with a longitudinal extent equal to the mode width of 1.7 mm/s, but without transverse selection.
The resulting atom number distribution is plotted in Fig.~\ref{fig1}(e) for various modulation amplitudes $A$, corresponding to different mean atom numbers. 
Tracing over one mode of a two-mode squeezed state results in a thermal density matrix $\sum_{i=0}^{\infty} P_i (n) \lvert i \rangle \langle i \rvert$~\cite{yurke.1987.obtainement,perrier.2019.thermal} where the atom number distribution is given by
\begin{equation}\label{eqn:ThermalStateDistribuition}
    P_i \left(n\right)  =\dfrac{n^i}{\left(1+n\right)^{i+1}}
\end{equation} 
and $n$ is the mean atom number. The functional form of this distribution is left unchanged by non-unit quantum efficiency with $n$ replaced by the mean {\it detected} atom number~\cite{perrier.2019.thermal}. In Fig~\ref{fig1}(e), $P_i$ is shown to agree  well with experimentally measured distributions. The thermal character of the state is corroborated by Fig.~\ref{fig1}(d) showing that the value of $g^{(2)}_{a}(\delta v_z=0)$ approaches~2~\cite{perrier.2019.thermal}.

\textit{Entanglement---}Two-mode entanglement in bosonic systems can be inferred from number correlations, provided the underlying quantum state remains Gaussian~\cite{gondret.2025.quantifying}. Within the Bogoliubov framework, the condensate is treated as a classical background that serves as a particle reservoir for collective excitations. The Hamiltonian is then quadratic, describing non-interacting quasiparticles and thus preserving the Gaussianity of the initial thermal states~\cite{robertson.2017.controlling}. This approximation remains valid as long as the condensate depletion is small. The data in Fig.~\ref{fig1}(b) confirms the small depletion.

\begin{figure}
    \centering
    \includegraphics[width=3.375in]{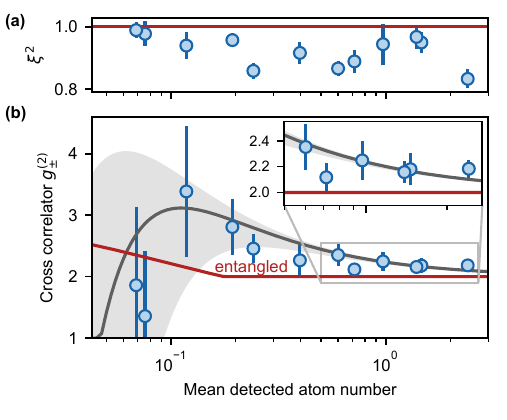}
    \caption{Normalized variance
    (a) and two-body correlator (b) as a function of the mean \emph{detected} atom number, for modulation amplitudes between $A=3$ and $28\%$.
    The hold time $\Delta t$ was fixed at 1.6 ms (3 breathing periods).
    The red line indicates unity (a) and the threshold for the entanglement witness of Ref.~\cite{gondret.2025.quantifying} in panel (b), assuming a quantum efficiency of 25\%.
    The gray curve shows the expected value assuming a two-mode squeezed thermal state with an initial temperature of 25(5) nK and a quantum efficiency of 25\%. 
    The width of the gray band reflects the uncertainty in the temperature.
    Error bars denote one standard deviation uncertainty and are computed using a bootstrap analysis.
    The value of $\xi^2$ was not corrected for the quantum efficiency.
    }
    \label{fig2}
\end{figure}

The normalized two-body cross-correlator of a zero-mean Gaussian state can be expanded using Wick’s theorem~\cite{Kardar_2007}, yielding
\begin{equation}\label{eq:g2_def}
g^{(2)}_{\pm} =\frac{\langle\hat{a}^\dagger_{+}\hat{a}^\dagger_{-}\hat{a}_{+}\hat{a}_{-} \rangle}{n_{+} n_{-}} = 1 + \frac{|\langle\hat{a}_{+}\hat{a}_{-}\rangle|^2}{n_{+} n_{-}} + \frac{|\langle\hat{a}_{+}\hat{a}_{-}^\dagger\rangle|^2}{n_{+} n_{-}},
\end{equation}
where $\hat{a}_\pm$ are the annihilation operators for the positive and negative momentum sideband modes, and $n_\pm = \langle \hat{a}_\pm^\dagger \hat{a}_\pm \rangle$ their mean occupations. In our experiment, the thermal statistics exhibited by each mode imply that they have zero mean ($\langle \hat{a}_\pm \rangle = 0$), and are not single-mode squeezed ($\langle \hat{a}_\pm^2 \rangle = 0$)~\cite{avagyan.2023.multimode}, validating the expansion of Eq.~(\ref{eq:g2_def}). For these Gaussian states, Ref.~\cite{gondret.2025.quantifying} shows that two-mode entanglement is certified whenever $g^{(2)}_\pm$ exceeds a population-dependent threshold, shown in red in Fig.~\ref{fig2}(b).

The measured
$g^{(2)}_\pm$ is 
shown in Fig.~\ref{fig2}(b) as a function of the mean detected atom number. 
This is done by varying the modulation amplitude $A$ between 3 to 28\%, fixing the hold time to 3 breathing periods. The value of  $g^{(2)}_\pm$ is given by the mean value of the product of the occupation numbers in two opposite voxels, normalized by the product of their means. Ideally, the value of $g^{(2)}_\pm$ should be computed in the limit of vanishing voxel size~\cite{gomes.2006.theory}. Here, $g^{(2)}_\pm$ is computed between voxels with a transverse size $\Delta v_\perp$, which matches the fitted full width at half maximum of the transverse density profile, and a longitudinal size of $\Delta v_z=1.4$~mm/s. This size is smaller than the correlation length to mainly pick out a single mode, yet large enough to ensure sufficient signal-to-noise.  
On the horizontal axis, the plotted population refers to the average number of detected atoms in the two sideband voxels~\footnote{The population is rescaled by the factor $\prod_i \text{erf}(\Delta v_i/ 2\sqrt{2} \sigma_i)$ (with $i = x, y, z$) to account for the fact that the voxels do not include all the atoms in each mode.}. Assuming that the two-mode state is Gaussian, we see that the two modes are entangled for a large range of mean detected atom numbers.
Since the trap opening is approximately adiabatic~\cite{gondret.2025.parametric}, this entanglement is inherited from the quasiparticles.
Thus, if the modes were initially unentangled, it demonstrates successful creation of quasiparticles from vacuum.

\textit{Two-mode squeezing model---}We now discuss additional checks that the system exhibits the phenomenology expected in our analog preheating picture. First, we compare the observed correlations with those predicted for parametric pair creation in a homogeneous background from modes initially in thermal states. 
Within the Bogoliubov approximation, the two-mode state remains Gaussian, and we have $\langle \hat{a}_{+} \hat{a}_{-}^\dagger \rangle =0$
on account of the homogeneity~\cite{robertson.2017.controlling}.
The two-mode state is fully described by the sideband populations $n_\pm$ and the anomalous correlator $|\langle \hat{a}_+ \hat{a}_- \rangle|$, with~\cite{busch.2014.quantum}
\begin{equation}\label{eq:tmsth}
    \begin{split}
        n_\pm &= (  \nth + 1/2 ) \cosh(2r) - 1/2,\\
        \lvert \langle \hat{a}_+ \hat{a}_- \rangle \rvert &= ( \nth + 1/2 ) \sinh(2r),
    \end{split}
\end{equation} 
where $r$ is the squeezing parameter~\cite{Martin:2021znx} and $\nth = 1 / (e^{\hbar\omega_\perp / k_B T} - 1)$ is the initial thermal occupation of each mode at energy $\hbar \omega_\perp$.
Entanglement occurs when $\lvert\langle \hat{a}_+ \hat{a}_- \rangle \rvert^2 > n_+ n_-$~\cite{robertson.2017.assessing}.
Since $\langle \hat{a}_{+} \hat{a}_{-}^\dagger \rangle =0$ here, by Eq.~\eqref{eq:g2_def} this is equivalent to $g^{(2)}_\pm > 2$ regardless of the mode occupation (i.e., in this model the equivalent of the red line in Fig.~\ref{fig2}(b) is a horizontal line at the value 2). 
The measured temperature 25(5)~nK corresponds to $\nth=0.18(8)$, or to a detected population of $0.04(2)$. The parametric growth of Eq.~\eqref{eq:tmsth} is seeded by both thermal and quantum fluctuations, whose contributions are respectively $\nth=0.18(8)$ and 1/2, but only the $1/2$ terms encoding vacuum fluctuations allows $\vert \langle \hat{a}_+ \hat{a}_- \rangle \vert^2 > n_+n_-$, signaling entanglement~\cite{micheli.2024.dissipative}. The low temperature is key to detecting entanglement using our witness. For larger $\nth$ we could still detect entanglement in principle but only with a larger gain and lower maximal value of $g^{(2)}_\pm$ which would be harder to distinguish from the value 2. In Fig.~\ref{fig2}(b), the gray band shows the theoretical prediction of this model, assuming a detection efficiency of 25\% and including uncertainty in the temperature. Measured correlations fall within this area and are thus compatible with the predictions of the model.

Second, the normalized variance $\xi^2 = \mathrm{var}(n_+ - n_-)/(n_+ + n_-)$ provides a slightly different and complementary way to 
characterize the state of the produced quasiparticles.
The variance is a non-classical signature which does not depend on the Gaussianity of the state
nor on its two-mode nature. Thus, we improve the signal-to-noise ratio by choosing large voxels centered on each sideband~\cite{jaskula.2010.subpoissonian, kheruntsyan.2012.violation}.
The results are shown in Fig.~\ref{fig2}(a).
At all but the lowest mode populations, the measured values of $\xi^2$ fall below shot noise level $\xi^2=1$, 
confirming the non-classical nature of the state~\footnote{Since the mode populations are equal, a normalized variance below unity directly signals a violation of the classical Cauchy–Schwarz inequality~\cite{finke.2016.nonclassical}.
Assuming that the particles are identical bosons, this violation also reveals \emph{particle} entanglement between the quasiparticles~\cite{wasak.2014.cauchy}.}.

\begin{figure}
    \centering
    \includegraphics[width=3.375in]{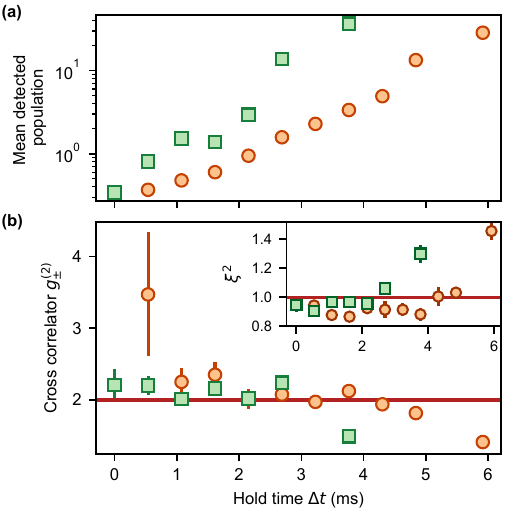}
    \caption{Mean detected population (a) and cross-correlation (b) as a function of the hold time $\Delta t$.
    Data is shown for two different modulation depths (orange circles 18\%, green squares 25\%).
    The inset shows the normalized variance. At late times, entanglement cannot be inferred. Error bars denote one standard deviation computed using a bootstrap analysis.
}
\label{fig3}
\end{figure}

\textit{Time evolution---}We investigate the time evolution by fixing the modulation amplitude and varying the hold time. The results are shown in Fig.~\ref{fig3} for amplitudes of 18\% (orange circles) and 25\% (green squares).  The mean quasiparticle population, shown in Fig.~\ref{fig3}(a), exhibits exponential growth, consistent with expectations from parametric resonance~\cite{micheli.2022.phonon} and our two-mode squeezing model, see Eq.~(\ref{eq:tmsth}) and Ref.~\cite{gondret.2025.parametric}. The evolution of $g_{\pm}^{(2)}$ and $\xi^{2}$ is shown in Fig.~\ref{fig3}(b). As the hold time $\Delta t$ increases, $g^{(2)}_\pm$ gradually approaches the value 2. This behavior is easily understood: although the anomalous correlation $|\langle \hat{a}_+ \hat{a}_- \rangle|$ increases with time, the ratio $|\langle \hat{a}_+ \hat{a}_- \rangle|^2 / (n_+ n_-)$ decreases as the population grows.  On the other hand, for hold times larger than 3 ms, the correlator begins to fall below 2 while the normalized number variance exceeds unity, a behavior not captured by Eq.~\eqref{eq:tmsth} nor its dissipative extension~\cite{micheli.2022.phonon,micheli.2024.dissipative}.  In this regime, we cannot infer entanglement. The onset of this apparent loss of entanglement for the two modulation amplitudes occurs for similar mode populations, consistent with the behavior seen in Ref.~\cite{robertson.2018.nonlinearities} when considering quasiparticle interactions.

\textit{Conclusion---}We have observed entanglement between parametrically excited quasiparticle modes of a quantum fluid. This entanglement demonstrates the role of quantum fluctuations in seeding the parametric growth, in analogy with cosmological preheating. While it has its limitations~\cite{Note1}, the relevance of the analogy is strengthened by the late-time dynamics. As noted above, the growing quasiparticle population leads to an interaction-dominated regime, of which the apparent loss of entanglement in Fig.~\ref{fig3} may be an indicator. This regime is characterized by a rich phenomenology, including decoherence of the resonant modes~\cite{robertson.2018.nonlinearities}, 
secondary peaks as observed in similar hydrodynamic experiments~\cite{gregory.2025.tracking}, and loss of Gaussianity~\cite{schweigler_decay_2021, bureik_suppression_2025}. These phenomena can be seen as steps towards thermalization, and taken together they are analogous to 
cosmological reheating~\cite{chatrchyan.2021.analog}. In addition, as the BEC keeps breathing, the energy and coherence of the driving field will be lost due to backreaction from the produced quasiparticles, the effect of vacuum fluctuations playing again a crucial role~\cite{Butera:2022kwi}. Future work will investigate these effects. \\

\acknowledgments
\textit{Acknowledgments}---We thank David Clément and Nicolas Pavloff for their careful reading of the manuscript and Maxime Jacquet for fruitful discussions.  A. M. thanks Ryo Namba for discussions on preheating, and is grateful to all iTHEMS members, in particular Tsukasa Tada, for providing a supportive research environment. The research leading to these results has received funding from QuantERA Grant No. ANR-22-QUA2-000801 (MENTA), ANR Grant No. 20-CE-47-0001-01 (COSQUA), and the LabEx PALM (ANR-10-LABX-0039PALM). 
V.G. acknowledges funding from the Conseil Régional, Île-de-France (DIM SIRTEQ) and the Quantum Saclay program FQPS (ANR-21-CMAQ-0002). 
R.D. acknowledges a PhD grant ref. 2024.03181.BD from the Portuguese Foundation for Science and Technology (FCT).
S.R. is funded by the CNRS Chair in Physical Hydrodynamics (ANR-22-CPJ2-0039-01).
\\

\textit{Data availability---}The data that support the findings of this manuscript are available in a public Zenodo repository at~\cite{gondret.2025.zenodo}.

\bibliography{dynamical_casimir}
\appendix
\onecolumngrid
\begin{center}
\large\textbf{End Matter}
\end{center}
\twocolumngrid

\textit{Method}---The parametric generation of correlated quasiparticles was previously reported by our group~\cite{jaskula.2012.acoustic}, but we were unable to observe non-classical effects. Here, we summarize the experimental improvements--further detailed in Ref.~\cite{gondr2025}--that enabled the detection of entanglement.
We have added a horizontal laser beam to produce a crossed dipole trap and used a 5~W instead of a 1.5~W laser. This has improved stability and added parameters to control the evaporative cooling inside the dipole trap. The shot-to-shot fluctuations in the BEC arrival time and hence of the vertical velocity of the atoms have been reduced by almost a factor 10. These fluctuations are now small enough to not degrade the measured value of the two-body cross-correlator. We have also been able to reach lower temperature by fine tuning the evaporation procedure.
The second improvement is the protocol to produce the parametric amplification: we now excite the breathing mode which we deliberately avoided in Ref.~\cite{jaskula.2012.acoustic}. This transverse excitation minimizes heating during the process~\cite{chevy.2002.transverse, jackson.accidental.2002}. In addition, we kept the modulation and the hold time short~\cite{robertson.2017.controlling, pylak.2018.influence}. Both of these allowed us to remain within a regime of low quasiparticle population where (i) 
their mutual interactions are negligible and their residual decay rate is small~\cite{micheli.2022.phonon, micheli.2024.dissipative, gondret.2025.parametric, robertson.2018.nonlinearities} and (ii) the visibility of entanglement is optimal~\cite{robertson.2018.nonlinearities}.
Finally, to ensure proper detection of both sidebands and to avoid detector saturation, we employ a pulse-shaped Bragg deflector, recently developed in our group~\cite{leprince.2025.coherent}.\newline
The data reported in this work represents approximately $70\,000$ experimental realizations, collected with good stability over a three-week period.\\

\textit{Detector quantum efficiency}---An accurate estimate of the quantum efficiency of the MCP detector is difficult. 
Earlier work on similar detectors on metastable helium yielded values ranging from 20\% to 50\%~\cite{lopes_atomic_2015,tenart_observation_2021,Kannan:2024}. For plotting purposes, we take 25(5)\%, compatible with the estimation made on our own detector. In the main text we compare our experimental results to two models. In both we assume that we produce two-mode Gaussian states with thermal marginal distribution, see Eq.\eqref{eqn:ThermalStateDistribuition}, meaning that $\langle \hat{a}_j\rangle=\langle \hat{a}^2_j\rangle=0$ for $j=\pm$. While in the two-mode squeezing model governed by Eq.~\eqref{eq:tmsth} we assume that $\langle\hat{a}_+^\dagger\hat{a}_-\rangle=0$, the one of Ref.\cite{ gondret.2025.quantifying} does not. The entanglement witness thus differs and we also need to look at the role of the detector non-unit quantum efficiency in these witnesses.

As pointed out in the main text, the entanglement witness of the first model is  $g^{(2)}_\pm>2$. Since the cross-correlator is a ratio, it does not depend on the quantum efficiency.\newline
In the second model, we have shown in Ref.\cite{ gondret.2025.quantifying}, that witnessing $g^{(2)}_\pm>g_E^{(2)}$ implies entanglement, where 
\begin{equation}\label{eq:g2_non_sep}
    \begin{split}
      \text{if } n_+n_- < 1/2,\, \, \, \, &   g^{(2)}_{E} =2 +\frac{1/2 -n_+n_-}{2n_-n_++ + n_- + n_+ +1/2},\\
      \text{if } n_+n_- \geq 1/2,\, \, \, \, &   g^{(2)}_{E} =2. \\
    \end{split}
\end{equation}
Here the witness depends on the mean populations and hence on the quantum efficiency $\eta$. They can be inferred 
from the detected mean population $n_j^{\rm(det)}$ as $n_j=n_j^{\rm(det)}/\eta$. Thus, at fixed mean detected population, the smaller the efficiency, the smaller the entanglement threshold.  In Fig.~\ref{fig2}, a 25\% detection efficiency is assumed.

A higher quantum efficiency would shift the entanglement witness curve of Fig.~\ref{fig2} to the right but even a quantum efficiency of 100\% would not change our conclusion on entanglement.\\

\begin{figure}
    \centering
    \includegraphics[width=3.375in]{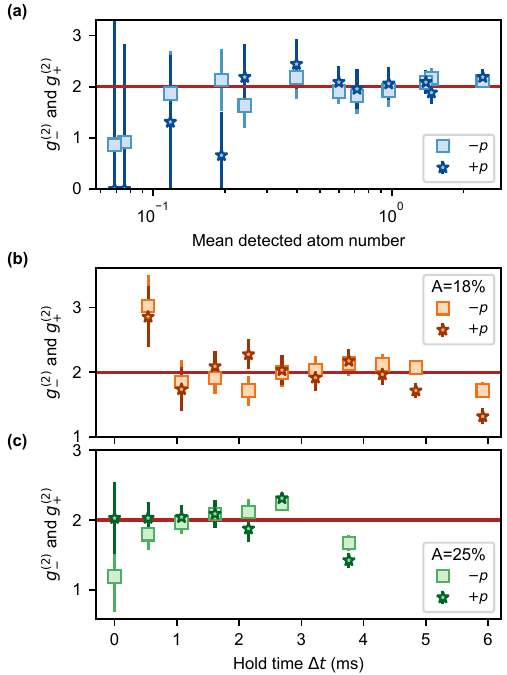}
    \caption{(a) The auto-correlator computed with the same dataset as in Fig.~\ref{fig2}, in which we vary the modulation depth for a constant hold time.
    Panels (b) and (c) show the corresponding auto-correlator for the dataset of Fig.~\ref{fig3} where the hold time is varied with a modulation depth of respectively 18\% and 25\%.
    The light square markers show the negative velocity sideband and the dark stars the positive velocity one.
    The error bars are computed from the expected error for a thermal law with the same detected atom number.}
    \label{fig4}
\end{figure}

\textit{Auto-correlations}---For purposes of comparison, we show in Fig.~\ref{fig4} the normalized two-body auto-correlators for the same data as that we used for Figs.~\ref{fig2} and \ref{fig3}.
The normalized two-body auto-correlators are defined as
\begin{equation}
    g^{(2)}_{j}=\frac{G^{(2)}_j}{n_j^2}=\frac{\langle(\hat{a}_j^\dagger)^2\hat{a}_j^2\rangle}{n_j^2}
\end{equation}
where $j=+$ or $-$. Note that $j\neq \pm$ as we look at auto-correlator and not cross-correlators. These auto-correlators are always consistent with the value 2 except in the case of very weak or very strong excitation. For weak excitation, the number of atoms in the sidebands is very low, resulting in large error bars, that we discuss below. In addition, the signal in the voxels is contaminated by other residual atoms. In the case of strong excitation, the decrease in the correlation may indicate a non-Gaussian state. 

The error bars shown in Fig.~\ref{fig4} are not obtained using bootstrap techniques, but are instead estimated based on the number of repetitions $N_{\text{shot}}$ and the mean detected atom number in the voxel $n_j$. When the population is sufficiently large, this estimation yields results consistent with a bootstrap analysis. However, when the mean detected atom number is very low, the available statistics are insufficient to reliably estimate error bars. In some cases, for instance, no doublets are recorded at all, due to limited statistics. Thus, the error bars on $g^{(d)}_j$ are given by 
\begin{equation}
    \Delta g^{(d)}_j=\frac{d!}{\sqrt{N_{\text{shot}}}}\sqrt{\frac{\text{Var}[(\ad)^d\an^d]}{[(d!)n_j^d]^2}+d\frac{\text{Var}[\ad\an]}{n_j^2}}\label{eq:gsmalld}
\end{equation}
where the variance is given by 
\begin{equation}\label{eq:variance}
    \text{Var}[(\ad)^d\an^d] =  \sum_{k=0}^{d}k!\binom{d}{k}^2 (2d-k)!n_j^{2d-k} - \left[d!n_j^{d}\right]^2.
\end{equation}
and only depends on the mean detected atom number $n_j$ and the number of experimental cycles $N_{\text{shot}}$.
In Eq.~\eqref{eq:gsmalld}, $d!$ represents the value of the normalized $d$-body autocorrelator $g^{(d)}_j$ for a thermal statistics.

In the following, we derive Eq.~(\ref{eq:variance}) for a state with a thermal statistics. To do so, we need to evaluate
\begin{equation}\label{eq:variance1}
    \text{Var}[(\ad)^d\an^d] = \braket{(\ad)^d\an^d(\ad)^d\an^d} - \braket{(\ad)^d\an^d}^2 \, 
\end{equation}
The second term in Eq.~(\ref{eq:variance1}) is the $d$-body auto-correlator which is given by $ G^{(d)}_j = \braket{(\ad)^d\an^d} =(d!)\, n_j^d$ in the case of a thermal statistics~\cite{mandel.1995.optical}.
To evaluate the first term, we normal order the creation and annihilation operators as~\cite{blasiak.2007.combinatorics}
\begin{widetext}

\begin{equation}
\begin{split}
(\ad)^d\an^d(\ad)^d\an^d  &
 = \mathopen{:}\underbrace{(\ad)^d\an^d(\ad)^d\an^d }_{\text{no pair removed}}\mathclose{:}\,  + :\underbrace{(\ad)^d\an^{d-1}(\ad)^{d-1}\an^d }_{\text{remove 1 pair}}: \times \underbrace{\binom{d}{1}\times \binom{d}{1} }_{\text{\# of pairs}}\\ 
&\quad + :\underbrace{(\ad)^d\an^{d-2}(\ad)^{d-2}\an^d }_{\text{remove 2 pairs}}:\times \underbrace{ \binom{d}{2}\times \binom{d}{2} \times2}_{\text{choose }\hat{a} \text{ then }\hat{a}^\dagger\text{, pair them}} + :\underbrace{(\ad)^d\an^{d-3}(\ad)^{d-3}\an^d }_{\text{remove 3 pairs}}:\times \underbrace{ \binom{d}{3}\times \binom{d}{3} \times 3\times 2}_{\text{choose }\hat{a} \text{ then }\hat{a}^\dagger\text{, pair them}}+...
\end{split}
\end{equation}
\end{widetext}
from which we obtain
\begin{equation}
    \braket{(\ad)^d\an^d(\ad)^d\an^d} = \sum_{k=0}^{d}k!\binom{d}{k}^2G^{(2d-k)}_j.
\end{equation}

\end{document}